\documentclass[aps,prl,showpacs,reprint,superscriptaddress,altaffilletter,altaffillsymbol,amssymb,amsmath]{revtex4-1}

\newcommand {\etal}{{\it et al.}}

\newcommand {\ryx}{\rho_{yx}}
\newcommand {\rn}{\rho_{yx}^{\rm{N}}}
\newcommand {\ra}{\rho_{yx}^{\rm{A}}}
\newcommand {\rt}{\rho_{yx}^{\rm{T}}}
\newcommand {\Dryx}{\Delta\rho_{yx}}
\newcommand {\DrTyx}{\Delta\rho^T_{yx}}
\newcommand {\um}{\mathrm{\mu m}}

\newcommand {\Beff}{B _{\mathrm{eff}}}
\newcommand {\ask}{a _{\mathrm{sk}}}

\newcommand {\Q}{{\bm Q}}

\newcommand {\n}{{\bm n}}

\newcommand {\Hc}{H_{\mathrm{c}}}
\newcommand {\Hperp}{H_{\perp}}
\newcommand {\Rs}{R_{\mathrm{s}}}
\newcommand {\Ms}{M_{\mathrm{s}}}

\usepackage{graphicx}
\usepackage{bm}
\usepackage{pifont}
\usepackage{amsmath}

\begin{document}

\title{Discretized Topological Hall Effect Emerging from Skyrmions in Constricted Geometry}

\author{N. Kanazawa}
\affiliation{Department of Applied Physics and Quantum Phase Electronics Center (QPEC), University of Tokyo, Tokyo 113-8656, Japan}
\author{M. Kubota}
\altaffiliation[Present Address: ]{Technology and Business Development Unit, Murata Manufacturing Co., Ltd., Nagaokakyo, Kyoto 617-8555, Japan.}
\affiliation{RIKEN Center for Emergent Matter Science (CEMS), Wako, 351-0198, Japan}
\affiliation{Power Electronics R\&D Division, ROHM Co., Ltd., Kyoto 615-8585, Japan}
\author{A. Tsukazaki}
\affiliation{Institute for Materials Research, Tohoku University, Sendai 980-8577, Japan}
\author{Y. Kozuka}
\affiliation{Department of Applied Physics and Quantum Phase Electronics Center (QPEC), University of Tokyo, Tokyo 113-8656, Japan}
\author{K. S. Takahashi} 
\affiliation{RIKEN Center for Emergent Matter Science (CEMS), Wako, 351-0198, Japan}
\author{M. Kawasaki} 
\affiliation{Department of Applied Physics and Quantum Phase Electronics Center (QPEC), University of Tokyo, Tokyo 113-8656, Japan}
\affiliation{RIKEN Center for Emergent Matter Science (CEMS), Wako, 351-0198, Japan}
\author{M. Ichikawa}
\affiliation{Department of Applied Physics and Quantum Phase Electronics Center (QPEC), University of Tokyo, Tokyo 113-8656, Japan}
\author{F. Kagawa} 
\affiliation{RIKEN Center for Emergent Matter Science (CEMS), Wako, 351-0198, Japan}
\author{Y. Tokura} 
\affiliation{Department of Applied Physics and Quantum Phase Electronics Center (QPEC), University of Tokyo, Tokyo 113-8656, Japan}
\affiliation{RIKEN Center for Emergent Matter Science (CEMS), Wako, 351-0198, Japan}

\date{\today}

\begin{abstract}
We investigate the skyrmion formation process in nano-structured FeGe Hall-bar devices by measurements of topological Hall effect, which extracts the winding number of a spin texture as an emergent magnetic field. Step-wise profiles of topological Hall resistivity are observed in the course of varying the applied magnetic field, which arise from instantaneous changes in the magnetic nano-structure such as creation, annihilation, and jittering motion of skyrmions. The discrete changes in topological Hall resistivity demonstrate the quantized nature of emergent magnetic flux inherent in each skyrmion, which had been indistinguishable in many-skyrmion systems on a macroscopic scale.
\end{abstract}

\pacs{72.25.Ba, 75.25.-j, 85.75.-d}

\maketitle

Recent studies on condensed matters with peculiar band topology \cite{Fang,Geim,Hasan,Murakawa} have established that emergent magnetic field arises from the Berry curvature in the momentum space and leads to unique electromagnetic responses \cite{Berry,Xiao,Nagaosa}, such as quantized anomalous Hall effects in topological insulators \cite{Checkelsky1,Xue,Checkelsky2}. Even for originally non-topological band (momentum-space) structures, materials may acquire a topological nature when the electrons interact with a specific spin texture in real space. The magnetic skyrmion in chiral magnets is an intriguing example of such topological spin textures, in which constituent spins point in all directions wrapping a sphere \cite{Muhlbauer,Yu1}. Such a winding spin texture whose topology can be labeled as the integer winding number $-1$ exerts a quantized emergent magnetic flux of $-\phi_0=-h/e$ on conduction electrons in case of strong limit of spin-charge coupling; as a result, topological Hall effect (THE), besides normal and anomalous Hall effects (NHE and AHE), can be observed as a hallmark of the skyrmion formation \cite{Neubauer,Lee1,Kanazawa,Ritz,Lee2,Franz}.

In addition to the topological aspect, the skyrmion possesses the small-sized particle nature (its typical diameter is 3-200 nm), which potentially enables magnetic non-volatile memories with ultrahigh density \cite{Fert,Tokura}. To fully benefit from such nano-sized characteristics, a methodology for identifying individual skyrmion needs to be developed. Although it has been demonstrated that observation and even manipulation of a single skyrmion are feasible by means of Lorentz transmission electron microscopy \cite{Yu1} and spin-polarized scanning tunneling microscopy \cite{Heinze,Romming}, identifying individual skyrmions by electrical means remains as a challenge for developing skyrmion-based magnetic memories. In this context, we can envisage that the quantized emergent magnetic flux inherent in the skyrmion may play a crucial role in identifying a single skyrmion. Most notably, smaller skyrmions for increasing recording density facilitate their detection with use of the emergent flux, contrary to conventional case where smaller magnetic bits are generally more difficult to detect: Effective magnetic field observed in Hall voltage increases inversely with the square of skyrmion size $\ask$, following the relation of $\Beff = -\phi_0/\ask^2$, which ideally reaches approximately 4000 T in a 1-nm skyrmion. Nevertheless, such a remarkable feature of skyrmions remains elusive, presumably because a large number of skyrmions involved in the macroscopic system studied so far inevitably smears out the quantized nature of the skyrmion \cite{comment1,comment3}.

In this Rapid Communication, we demonstrate discretized changes in the emergent magnetic flux arising from a finite number of skyrmions, by measuring topological Hall resistivity of nanoscale FeGe Hall-bar devices. By measuring the inclination-angle dependence of topological Hall resistivity, which works as a test for skyrmion formation in thin films \cite{Yokouchi}, we can ensure a hysteretic formation of skyrmions appearing as a hysteresis loop in Hall resistivity \cite{Chien}. In the devices with circuit line width between 50-250 nm, the discretized profiles of topological Hall resistivities show up in the hysteresis loops of Hall resistivities where rapid conversions between skyrmions with opposite core-magnetizations occur. Further miniaturization to 32-nm-wide circuit removes the typical signature of skyrmion formation (a hysteretic topological Hall resitivity), which indicates that skyrmions are not able to form in a smaller area than half the size of themselves.

Hall-bar devices of various circuit-line-widths (32 nm - 250 nm and 10 $\um$) were fabricated from 40-nm-thick FeGe epitaxial thin film by using electron-beam lithography method. A top-view image of a 50-nm-wide device is examplified in Fig. 1(a). The electrical leads for current and voltage were made to have the same width ($w$); skyrmion formations in the intersectional areas of $w\times w$ were probed by measurements of the topological Hall effect \cite{supple}. The FeGe film was grown on a highly-resistive Si (111) substrate ($\rho > 1000$ $\mathrm{\Omega}$ cm) by codeposition of Fe and Ge at a substrate temperature of 325 $^\circ$C by molecular beam epitaxy. A very low concentration of impurity phases was verified by high-angle X-ray diffraction (XRD) as detailed in Supplemental Material \cite{supple}. Here we note that the electrical conduction at 2 K is not the ballistic one because the residual resistivity ranges between 50-150 $\mathrm{\mu\Omega}$ cm \cite{supple}. We do not have to take mesoscopic effect into account.

We first confirm skyrmion formation in our FeGe thin film by THE. Topological Hall resistivity $\rt$ associated with formation of a non-coplanar magnetic structure is usually defined as an additional component to the terms varying linearly with magnetic field $H$ and magnetization $M$, i.e., the normal and anomalous Hall resistivity $\rn$ and $\ra$, respectively. However, those two components ($\rn$ and $\ra$) may show $H$- and $M$-nonlinear dependence especially in multi-band metals \cite{Yokouchi,Taguchi}. 
Here we adopt another precise method for evaluating THE based on the $H$-direction-sensitive stability of skyrmions in thin films \cite{Yokouchi}. According to neutron scattering experiments on skyrmionic materials \cite{Muhlbauer}, the plane of skyrmion lattice, where the three magnetic modulation vectors $\Q$ lie, is strictly perpendicular to the magnetic field. Namely, energy gain from interactions among the spins on the same plane perpendicular to $H$ is crucial for stabilization of skyrmions. When $H$ is tilted from the normal vector of the thin film plane $\n$, $\Q$s have out-of-plane components, which means that spins near the surfaces lose their partner spins to interact with [Fig. 1(b)]. The skyrmion state is thus easily destabilized with $H$ tilted from $\n$ in thin films, and this feature should become more significant in thinner films compared to their skyrmion sizes.

Figure 1(c) shows magnetic-field dependence of Hall resistivity at various inclination angles $\theta$ at 2 K in a 10-$\um$-wide Hall-bar device. We observe a loop in Hall resistivity at $\theta=0^\circ$, which is recognized to be induced by residual skyrmions of metastable state dependent on magnetic-field history, as reported by Huang and Chien \cite{Chien}. The hysteresis loop dramatically shrinks with a slight change in the inclination angle: the loop size becomes less than the half just tilting at $\theta=4^\circ$ and nearly vanishes above $\theta=10^\circ$. This sudden reduction of the $\ryx$ loop, which is highly unlikely for the magnetization-related anomalous Hall effect in ordinary ferromagnets, represents the destruction of two-dimensional skyrmion structures as expected.

We extract topological Hall component in Fig. 1(d) by subtracting $\ryx$ at a high angle of $\theta=30^\circ$ consisting of normal and anomalous Hall components from low-angle data \cite{supple}. If the vertical component of $H$ to the film plane ($\Hperp$) is lower than the critical field $\Hc$, where the ferromagnetic state is induced, the difference between Hall resistivities at low and high ($\theta=30^\circ$) angles reads
\begin{eqnarray}
&& \Delta\ryx=\ryx(\theta)-\ryx(30^\circ)\notag\\
&& =[\rn(\theta)+\ra(\theta)+\rt(\theta)]-[\rn(30^\circ)+\ra(30^\circ)] \notag\\
&& =(R_0 B_{\perp} + R_{\mathrm{s}} M_{\perp} + R_0 \Beff \cos\theta) -  (R_0 B_{\perp} + R_{\mathrm{s}} M_{\perp}) \notag\\
&& = R_0 \Beff \cos\theta. \notag
\end{eqnarray}
Otherwise ($\Hperp>\Hc$), 
\begin{eqnarray}
&& \Delta\ryx=[\rn(\theta)+\ra(\theta)]-[\rn(30^\circ)+\ra(30^\circ)] \notag\\
&& =(R_0 B_{\perp} + \Rs \Ms\cos\theta) -  (R_0 B_{\perp} + \Rs \Ms\cos30^\circ) \notag\\
&& = \Rs\Ms (\cos\theta-\cos30^\circ). \notag
\end{eqnarray}
Here $R_0$ and $\Rs$ are normal and anomalous Hall coefficients, and $M_{\perp}$ and $\Ms$ the vertical component of magnetization $M$ and the saturated magnetization, respectively.
Although the extracted $\Delta\ryx$ inevitably includes anomalous Hall component $\Rs\Ms (\cos\theta-\cos30^\circ)$ at high fields, $\Delta\ryx$ for $\Hperp<\Hc$ is almost equivalent to $\rt$. The rapid shrinkage of $\rt$ is again highlighted in Fig. 1(d). As magnetization at the core of the skyrmion is anti-parallel to the field \cite{Muhlbauer,Yu1}, skyrmions with positive (negative) core-magnetization is stabilized at negative (positive) magnetic field. Hereafter, we refer to skyrmions with positive (negative) core-magnetization as core-up (down) skyrmions, which produce positive (negative) emergent magnetic fields. The hysteresis loop of $\rt$ demonstrates that core-up or core-down skyrmions are stabilized even at zero magnetic field depending on a magnetic-field history, as discussed later. Possibility of other interpretations of the hysteretic Hall resistivity is discussed in the Supplemental Material \cite{supple}; nevertheless, the skyrmion formation is the most plausible scenario to explain the hysteretic behavior.

\begin{figure}
\begin{center}
\includegraphics*[width=8cm]{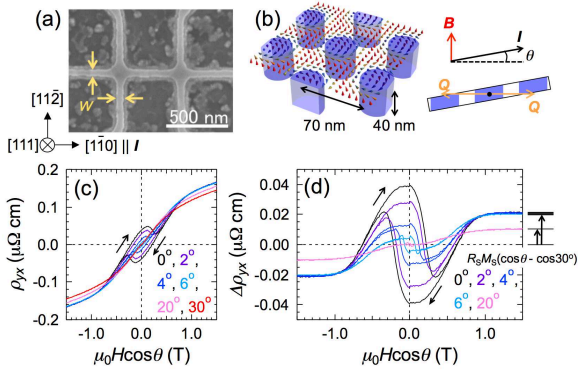}
\caption{(color online). (a) Scanning electron microscope image of a Hall-effect measurement circuit with line width of $w=50$ nm. (b) Schematic illustrations of the skyrmion-lattice state in a 40-nm-thick FeGe thin film. The magnetic modulation period in the FeGe thin film is assumed to be the same length of 70 nm as the bulk sample. Magnetic-field dependence of (c) Hall resistivity $\ryx$ and (d) $\Dryx$ at various inclination angles at 2 K in a 10-$\um$-wide circuit. (See text for the definition of $\Dryx$.) A sudden shrinkage of hysteresis loops in panels (c) and (d) represents declines in skyrmion numbers.}
\end{center}
\end{figure}

To pursue the quantized nature of emergent flux, we fabricated Hall-bar micro-devices and found that such a hysteresis loop representing skyrmion formation remain robust down to a 50-nm-wide device. We show the line-width dependence of Hall resistivity in Fig. 2. Devices with wider circuit lines than 50 nm show the characteristic hysteresis loops representing skyrmion formation \cite{comment4}. By contrast, Hall resistivity in a 32-nm-wide circuit almost traces the magnetization curve \cite{supple}. The collapse of the hysteresis loop with narrowing the circuit line represents that the critical size for the formation of skyrmions of FeGe lies between 32 nm and 50 nm \cite{comment2}; this critical size is to be compared with the helimagnetic period ($\approx70$ nm) or the lattice constant ($\approx 80$ nm) of skyrmion crystal of FeGe bulk crystal. We assume that a helix-like magnetic structure with no topological charge is realized in the 32-nm device as theoretically indicated by Du \etal \cite{comment3}.

Taking a close look at the hysteresis loops in the nano-scale circuits, we notice that Hall resistivity shows discontinuous changes with field variation, which indicates the discrete change in emergent magnetic field. 
\begin{figure}
\begin{center}
\includegraphics*[width=8cm]{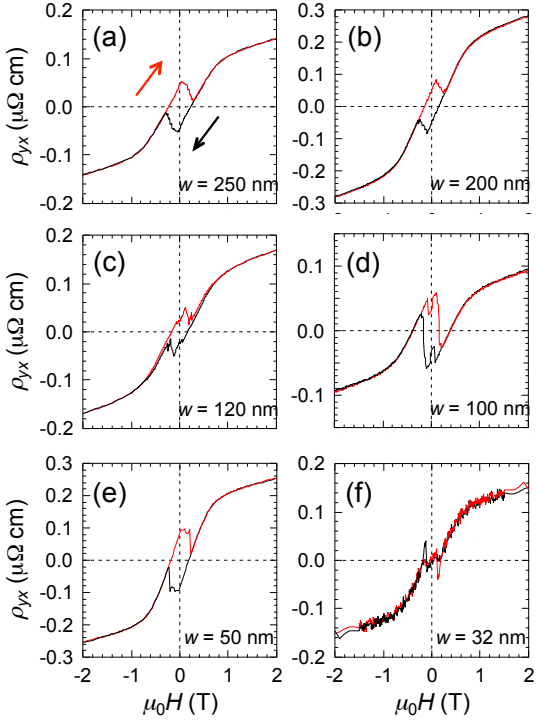}
\caption{(color online). (a)-(f) Magnetic-field dependence of Hall resistivity for FeGe nano-scale circuits with various widths ranging from 250 nm to 32 nm. We subtracted the $H$-symmetric component using the equation $\ryx (\pm H)=\pm\frac{[V_y(H)-V_y(-H)]t}{2I}$.}
\end{center}
\end{figure}
Figure 3 shows magnified drawings of hysteresis loops (insets) and difference between Hall resistivities in increasing and decreasing field processes, $\Delta\rt$ (main panels), in various-size devices; $\Delta\rt$ is plotted as the normalized value by its maximum value in the figure. Since the conventional Hall components, i.e., $\rn$ and $\ra$, do not show hysteresis behaviors (see Supplemental Material for the $M$-$H$ curve, which shows little hysteresis behavior \cite{supple}.), $\Delta\rt$ associated with the hysteresis loop represents the genuine contribution from the topological Hall effect. In contrast to the smooth variation of $\Delta\rt$ in the 10-$\um$ device [Fig. 3(a)], $\Delta\rt$ in nano-circuits of $w=50$-250 nm exhibit step-wise profiles [Figs. 3(b)-3(f)]. The noise level is one order of magnitude smaller than the smallest step height, 5 $\mathrm{n\Omega}$ cm, indicating that the discretized step can not be electrical noise of our experimental system \cite{supple}.
Constricted geometry comparable to the size of a few skyrmions emphasizes skyrmions' individuality, so that we successfully observed the quantized nature of emergent magnetic field as discrete changes in topological Hall resistivity mainly originating from creation/annihilation of skyrmions in changing the applied magnetic field.

Incidentally, we found that the step heights display considerable variation and are not equal to integer times of unit topological Hall resistivity $\rt=-R_0 \phi_0/w^2$, corresponding to creation/annihilation of one skyrmion. This indicates additional contributions to the discrete changes in emergent magnetic field, which we attribute to discontinuous motions of skyrmions via trapping by or releasing from impurity or defect sites in the course of field changes: this bears analogy to the Barkhausen effect, i.e., discontinuous changes in magnetization via the similar motion of ferromagnetic domains. The possible fluctuation of skyrmions' positions, especially skyrmions putting their feet on and off the verge of the probed square area of the Hall device, may produce ``halfway" discrete topological Hall effect. For examples, the fluctuation of $\rt$ around $\mu_0 H=0.2$-0.3 T in 120-nm-wide device [Fig. 3(d)] and dip structure of $\rt$ around $\mu_0 H=0$-0.1 T in 100-nm-wide device [Fig. 3(e)] can be assigned to discontinuous changes in skyrmion position. Here we note that the steady flow of skyrmions via the spin-transfer torque, which may also contribute to the variation in the step heights, probably does not occur in present study considering the enhanced critical current density in nano-structures \cite{supple,Iwasaki}. 

\begin{figure}
\begin{center}
\includegraphics*[width=8cm]{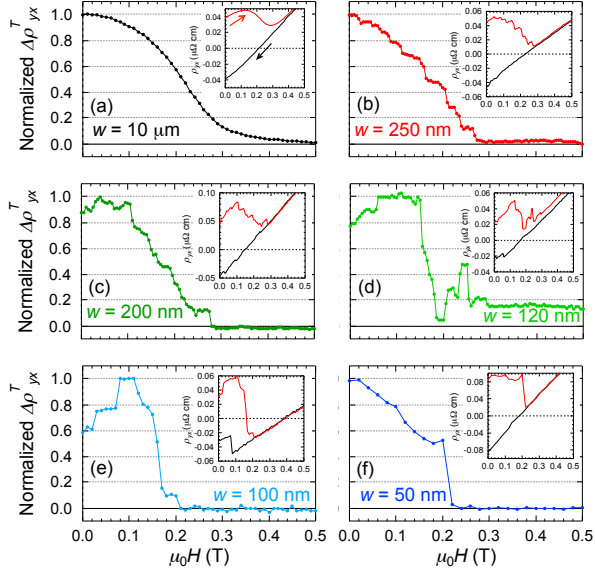}
\caption{(color online). (a)-(f) Magnetic-field dependence of difference between Hall resistiveities in increasing and decreasing field processes, which is denoted by $\Delta\rho^T_{yx}$. Normalized $\DrTyx$ by its maximum value is presented. Insets are magnified images of each $\ryx$ at low fields.}
\end{center}
\end{figure}

Hall resistivity data enable us to envisage the skyrmion formation process in the nano-structured Hall bars. Here we take, for example, the case of $w=250$ nm wide device, which can accommodate approximately 9 skyrmions at the overlap of current and voltage leads $w\times w$. Figures 4(a) and 4(b) are the conceivable development of numbers of core-up and -down skyrmions in increasing and decreasing field processes, respectively. In increasing magnetic field from a large negative field ($<-\Hc$) [Fig. 4(a)], the magnetic state is anticipated to trace the successive stages as illustrated in Fig. 4(e): first, core-up skyrmions gradually pop up [panel (ii) of Fig. 4(e)] in the background of ferromagnetic state of pointing-down moments (i); next, high-density core-up skyrmions, possibly skyrmion lattice state, is realized and remains around zero field as the metastable state (iii); as the magnetic field increase from zero to a positive value, core-up skyrmions are destabilized to rapidly disappear coalescing with neighboring ones (iv); and then form a convoluted helical structure (v); the helical structure fragments to form core-down skyrmions (vi), which then crystallize (vii) and subsequently begin to disappear (viii); finally, spins are fully polarized upward to form the ferromagnetic state (ix).

Magnetic-field profile of Hall resistivity can be simulated, as shown in Figs. 4(c) and 4(d), from the assumed skyrmion formation [Figs. 4(a) and 4(b)]. A unit of topological Hall resistivity produced by a single skyrmion is determined from $\Delta\rt$: Since the maximum value of $\Delta\rt$ represent the difference between topological Hall resistivities produced by close-packed core-up and -down skyrmions, the unit of $\rt$ is $\Delta\rt$ divided by twice the maximum capacity (i.e., $2\times9$ skyrmions) of the overlap area ($w\times w$). In the present example, the unit of $\rt$ is derived as approximately 5 $\mathrm{n\Omega\ cm}$ from $\Delta\rt/(2\times9)$. We show the simulated $\rt$ in Fig. 4(d) along with NHE and AHE evaluated from the slope of Hall resistivity at high fields above $\Hc$ and magnetization $M$, respectively. The overall profile of $\rt$ bears a remarkable resemblance to the extracted $\rt$ of the 10-$\um$ device [Fig. 1(d)], which supports our model. The total Hall resistivity as simulated [Fig. 4(c)] also shows a good agreement with the experimental results [Fig. 2(a)].  The discretized feature of $\rt$ especially shows up in parts of the hysteresis loop where the rapid switch between core-up and core-down skyrmions occurs, which we focus on in Fig. 3. 

\begin{figure}
\begin{center}
\includegraphics*[width=8cm]{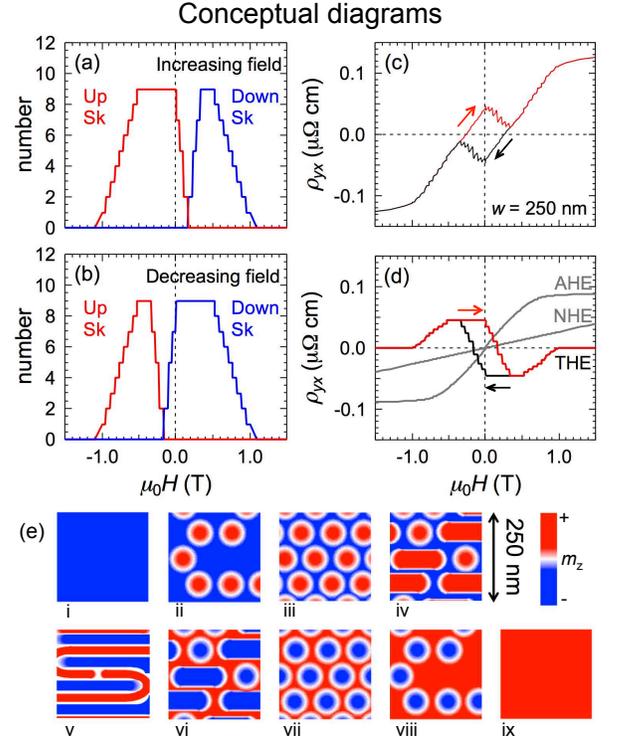}
\caption{(color online). Conceptual diagrams for skyrmion formation and corresponding Hall resistivities in a 250-nm-wide nano-scale circuit. Variations of numbers of core-up and core-down skyrmions as functions of magnetic field in (a) increasing and (b) decreasing field processes. Corresponding (c) Hall resistivity and (d) its decomposition into $\rn$ (NHE), $\ra$ (AHE), and $\rt$ (THE). (e) Series of schematic illustrations representing an expected development of magnetic structure as a function of magnetic field.}
\end{center}
\end{figure}

In conclusion, we observed the discretized topological Hall effect in nano-structured FeGe Hall-bar devices with their sizes of one or a few skyrmions. The discontinuous changes in topological Hall resistivity, arising from discontinuous motion, creation and annihilation of skyrmions, represent the quantized nature of the emergent magnetic field inherent in each skyrmion, which gets buried in many other skyrmions residing in a large-size sample used in prior experimental studies. The present study shows that the emergent magnetic field producing the topological Hall effect can be utilized to detect single skyrmion in future skyrmion-based nanodevices. Skyrmion formations in FeGe films and nano-devices require further validation by real-space observations, which are left as future challenges. 

The authors thank T. Arima, W. Koshibae, and T. Yokouchi for enlightening discussions. This work was supported by Young Scientists(Start-up) No. 26886005, Scientific Research(S) Nos. 24224009 and 24226002, and the Funding Program of World-Leading Innovative R\&D on Science and Technology (FIRST program) on ``Quantum Science on Strong Correlation" initiated by the Council for Science and Technology Policy, Japan.


\begin{thebibliography}{100}
\bibitem{comment4} All the hysteresis loop observed in the nano-devices shows the same magnetic-field history, i.e., positive (negative) topological Hall resistivity in the course of increasing (decreasing) field process. If we assume that some other nanoscale magnetic structure, such as deformed helical structures and ferromagnetic domains, which may show topological and hysteretic anomalous Hall effects, is the origin of the hysteresis loop, we cannot expect the unchanging magnetic-field history; those structures can equally produce both positive and negative signals.
\bibitem{comment2} Here we note that the unavoidable side damage in the etching process would have greater influence on the smaller devices, which might be a crucial damage in the 32-nm device related to the inhibition of skyrmion formation. However, considering the magnetization property of FeGe manifesting itself in the Hall resistivity in the 32-nm device, there should be some effects of miniaturization against skyrmion formation, although we could not estimate the exact value of the threshold size.
\end{thebibliography}
\end{document}